# High Level Programming for Heterogeneous Architectures

Oren Segal and Martin Margala
Department of Electrical and Computer Engineering
University of Massachusetts Lowell
Lowell, MA
{oren_segal,martin_margala}@uml.edu

Sai Rahul Chalamalasetti and Mitch Wright
HP Servers
Hewlett-Packard
Houston, TX
{sairahul.chalamalasetti,mitch.wright}@hp.com

*Abstract*—**This work presents an effort to bridge the gap between abstract high level programming and OpenCL by extending an existing high level Java programming framework (APARAPI), based on OpenCL, so that it can be used to program FPGAs at a high level of abstraction and increased ease of programmability. We run several real world algorithms to assess the performance of the framework on both a low end and a high end system. On the low end and high end systems respectively we observed up to 78-80 percent power reduction and 4.8X-5.3X speed increase running NBody simulation, as well as up to 65-80 percent power reduction and 6.2X-7X speed increase for a KMeans, MapReduce algorithm running on top of the Hadoop framework and APARAPI.**

## I. INTRODUCTION

Using GPUs to accelerate computation has been shown to be more power efficient then using CPUs alone [1]. Recent research is suggesting that using FPGAs as accelerators can further increase power efficiency for certain types of data-centric applications [2]. One main drawback to using FPGAs is the difficulty in programming them [3]. The traditional way to program FPGAs has been through the use of hardware description languages (HDLs) such as Verilog and VHDL; Languages which require technical abilities and know-how not found in the average computer programmer [6].

The emerging open programming standard for heterogeneous computing is OpenCL [4]. OpenCL offers a unified C programming model for any device that adheres to its standards. Recently Altera Corp released an OpenCL SDK for FPGAs [5] allowing FPGAs to be treated as OpenCL based accelerators. The OpenCL programming framework, although a major step in the effort to ease FPGA programming, still presents challenges to programmers since it requires both proficiency in C and in-depth understanding of the inner workings of the OpenCL programming model. In addition to that, recent surveys [6] show that a significant portion of software programmers use programming languages such as Java, C# and Python which offer rapid prototyping and easier software maintenance. Such programmers can not benefit from the computing power and energy efficiency that OpenCL, FPGAs and heterogeneous computing have to offer. In recent years several initiatives [10,11,18,19,29] have been launched in an effort to bring heterogeneous computing to the masses. One such effort is a framework called APARAPI [10] which was originally developed by AMD to target AMD GPU/APU architectures [12]. APARAPI is a Java based programming framework that aims to lower the heterogeneous programming bar. APARAPI offers programmers the ability to run heterogeneous computing applications using a familiar Java interface while hiding many of the intricate details that are associated with OpenCL programming. The main drawback of the APARAPI framework is that it was designed to specifically support AMD devices which make it unusable on other architectures. In this work our aim was to remove APARAPI's dependency on the AMD architecture essentially enabling it to run on any other heterogeneous architecture that is OpenCL compatible. Extend APARAPI to support FPGA based devices and investigate the use of APARAPI on real world parallel algorithms in the data center. In addition we explore the possibility of incorporating APARAPI-FPGA into the Hadoop [15] framework, an industry standard Java software framework used in massively parallel distributed processing of Big Data for collaborative filtering, clustering, and classification.

The rest of the paper is organized as follows. Section II gives an overview of the APARAPI framework. Section III reviews the steps performed in order to modify APARAPI to support Altera OpenCL for FPGA devices. Sections IV and V describe the software and hardware used in our experiments, Section VI details benchmark performance and power analysis of the modified framework on FPGAs, Section VII contains related work, Section VIII presents our conclusions and Section IX discusses future directions .

## II. OVERVIEW OF THE APARAPI FRAMEWORK

APARAPI is a Java based framework that allows a programmer to write code in high level Java and have this code automatically translated and divided into host code and OpenCL code. The user of APARAPI is freed from the details of querying, instantiating and transferring data to and from the OpenCL compatible device. Fig. 1 shows a typical OpenCL kernel written in Java, using the APARAPI framework.

As can be seen those few lines of code replace dozens of lines of code of traditional boiler plate OpenCL code necessary to get a simple OpenCL kernel running. In fact the original OpenCL version of this program written in C was over three hundred lines of code compared to seventy lines in the APARAPI version. That is more than 4X increase which in the world of software development translates to increased development effort and potential number of bugs [9]. The magic behind the simplification done by the APARAPI framework can be roughly divided into the following steps:
1. Compile the Kernel function into Java byte code
2. Locate the data members needed for data transfer between the host and the device
3. Generate an OpenCL kernel from the Java kernel function
4. Initiate the OpenCL subsystem and allocate the required memory buffers
5. Run the boiler plate code needed to setup the kernel, pass the parameters to the kernel and return the results.

Fig. 2 shows the APARAPI system model and the flow between the written code and the underlying OpenCL compatible hardware device.

```
final float inA[N] = .... // get a float array of data from somewhere
final float inB[N] = .... // get a float array of data from somewhere
final float result = new float[inA.length];

Kernel kernel = new Kernel(){
  @Override public void run(){
    int i= getGlobalId();
    result[i]=intA[i]+inB[i];
  }
};

Range range = Range.create(result.length);
kernel.execute(range);
```

Fig. 1. APARAPI Java code.

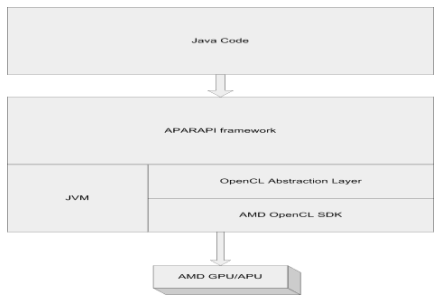

Fig. 2. Original APARAPI architecture

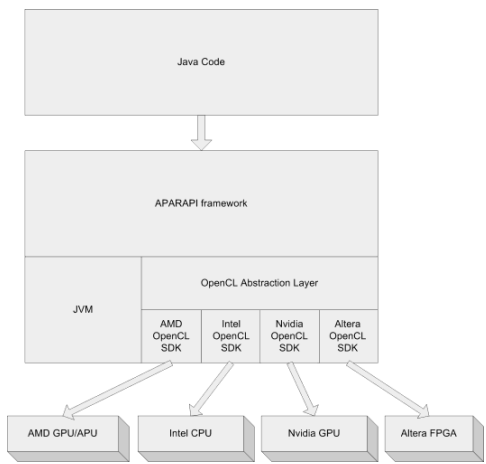

Fig. 3. Modified APARAPI architecture

## III. MODIFICATIONS TO THE APARAPI FRAMEWORK

Our work revolved around modifying the APARAPI framework so it can support other types of OpenCL compatible devices. Initially our efforts were geared towards targeting FPGAs, but we plan to investigate the ability to target multiple devices at the same time. Fig. 3 shows how the proposed multi target modified architecture would look like. In order to modify the framework to support Altera OpenCL for FPGAs the following steps were performed:
1. Disconnect the dependency on the AMD OpenCL SDK
2. Link the APARAPI code against the Altera OpenCL SDK
3. Make any code modifications needed to support the new types of devices, specifically FPGAs.

Step one and two required modification to the make files so they will be linked against the new set of Altera OpenCL libraries. Several libraries had to be replaced and minor modification changes made to accommodate that step. Step three was more involved and required some changes to the native APARAPI layer that calls into OpenCL. Common implementations of OpenCL (Nvidia, Intel, AMD etc.) all build OpenCL as a dynamic library potentially allowing a process to dynamically link several OpenCL versions and implementations. The Altera implementation needs to be statically linked to an executable. In addition Altera libraries were not thread safe which added an additional level of complexity, since we needed to add a layer to synchronize calls into the device from a single thread. In addition some minor deviations in the OpenCL API were detected and needed addressing. Host memory allocation was another area where modifications needed to be made. The Altera FPGA OpenCL API required 64 byte memory aligned allocations to function properly, but all in all working with an open standard such as OpenCL made the conversion possible in a matter of weeks. Memory allocation in Java for host code was another major hurdle. Unlike native code, written in C/C++, for example, in which memory allocation is controlled by the user, Java manages memory allocation and garbage collection for the user. This exacerbates the already existing memory bottleneck of acceleration devices. In fact to make acceleration feasible a minimum threshold of computation intensity must be reached [7]. A core requirement was to make sure that code that ran on the original APARAPI framework would run unmodified on the new version of the framework, allowing for code reuse and future merging of the two branches of the framework.

## IV. SOFTWARE LEVEL EXPERIMENT DESIGN

### A. Standalone Benchmark Algorithms

We first analyzed several standalone benchmarks that are part of the APARAPI distribution and compared the performance of an APARAPI application running on Java thread pool (CPU threads) and on an FPGA device. We chose three algorithms that can potentially exploit parallelism very well. The algorithms chosen were: Mandelbrot fractal set calculation, black-scholes option pricing and NBody physics simulation.
These algorithms were run on both APARAPI framework versions without any Java user code modifications. The OpenCL kernel that is auto generated by the APARAPI framework was fed to and automatically optimized by the Altera OpenCL compiler. Note that unlike with a traditional OpenCL compiler the Kernel compilation process on the Altera OpenCL compiler can take several hours since it has to generate the FPGA bit stream. The resulting synthesized Kernel is loaded on the fly to the FPGA when the Java user program is running.

*B. MapReduce Based Algorithms*

Since Big Data processing is a major and growing data center activity [36], we now turned our interest to MapReduce algorithms [14] based on the common Hadoop framework [15]. Specifically the K-Means algorithm [16] was chosen as a popular representative of a MapReduce algorithm. The Hadoop framework presented the following challenges for efficient use with an accelerator device:

1. It was not built to accelerate or take advantage of a single machine. Instead it relies on the massive use of multiple computational nodes.
2. The computation intensity of some of the algorithms is not sufficient to warrant exchanging data between the device and the host machine.

Because of the above care needs to be taken when choosing an algorithm and processing data to ascertain that the combination of data size and computation intensity per compute node reaches a critical mass.

*C. The K-Means Algorithm*

The K-Means algorithm is a clustering algorithm that attempts to group points around common centers. The input is a group of points and number of centers and the result is a list of centers and the points that belong to each center. Effectively one can use the algorithm to divide data into separate groups (clusters) and find their common center point.
The algorithm has three parameters:
N - The size of the data to be processed by the algorithm (number of points)
K – Number of centers on which the N points should be grouped around
D – Number of dimensions of each point

*D. K-Means Experiment Design*

Several previous attempts at accelerating the Hadoop framework involved using a modified MapReduce framework to accommodate for the limitations and strengths of accelerators [31, 32] in our experiment we attempted to explore a scenario where an accelerator is introduced to an existing unmodified Hadoop framework in a typical data center scenario and find what can be done within the confines of the existing framework. We started with a standard reference Java K-Means map reduce sequential algorithm for the Hadoop framework [38], converted it to work with floating number precision and optimized it. Note that standard Hadoop algorithms are sequential in nature i.e. not built to exploit single machine parallelism instead they rely on multiple machine parallelism to accelerate their operation. In the next stage we created a new parallel Java algorithm in APARAPI. An algorithm running on Hadoop is divided to two computational tasks: map and reduce. In order for the algorithm to run effectively on an FPGA we had to find the most computational intensive part of K-Means and make sure it has enough data to process on each invocation of the kernel. To do that we implemented the kernel as an aggregated mapping task, calculating the Euclidian distances of points from centers, which has the complexity of $O(NKD)$ floating point operations per Hadoop iteration. The APARAPI version of the algorithm can run in several modes, and we report the results for both JTP (Java Thread Pool) and FPGA mode (OpenCL on FPGA). Since FPGAs are versatile devices that can be optimized on a per algorithm basis the Altera OpenCL SDK allows a designer to specify, on a per OpenCL kernel basis, the level of loop unrolling(ULs) and number of parallel processing units(PUs) in the kernel to be synthesized on the FPGA. In addition to that the compilation tools allow an automatic optimization of the design in order to maximize performance (O3 option). We explore such different variations of the FPGA kernels and run the algorithm on different combinations of N, K and D to see the effects of the complexity and amount of data on the performance of different variations of the kernel.

V. HARDWARE LEVEL EXPERIMENT DESIGN

In order to simulate several typical data center scenarios we chose two types of systems:

1. Low end server equipped with an Intel i7 3770 3.4GHz processor with 12GB DDR3 1333MHz RAM, one that could be found in a typical ad hoc machine cluster.
2. High end server HP DL180 G6 with two Intel Xeon L5630 2.13GHz processors with 144GB DDR3 1333MHz RAM, one that would be more typically found in a traditionally organized data center environment.

Both machines were equipped with a Nallatech PCIe-385N A7 FPGA board (Fig 7.) with 8GB RAM connected through PCIE second generation connection to the host system. Power measurements were obtained using a wattsup pro power meter.

VI. RESULTS

*A. Standalone Benchmark Algorithms*

*1) Speed Analysis*

Performance results for the Mandelbrot fractal set calculation, black-scholes option pricing and NBody physics simulation are presented in Table II. All three algorithms show significant speed increase when compiled with auto optimization on. These algorithms are computational intensive and the overhead associated with data exchange over PCIE between the host and the FPGA device is relatively small, meaning that the ratio between the amount of computations done in the FPGA kernel and the amount of data that is transferred between the host machine and the FPGA in each kernel invocation is relatively high.

*2) Power Analysis*

As can be seen in Table I the mere inclusion of the FPGA in the system cause a 19.5W power increase in idle mode and running CPU threads (JTP) at max utilization causes an additional power drain that takes the total power overhead to 20.6W and 24.7W on systems A and B respectively. The power consumption balance moves in favor of the FPGA system when the system is at high utilization. When the system operates at 100 percent utilization in JTP mode, all

TABLE I.     AVG POWER CONSUMPTION

| System Type | System Power in Watts |
| --- | --- |

|  | Shutdown | Idle | JTP | FPGA |
|---|---|---|---|---|
| CPU & FPGA | 1.7 | 42.6 | 88.15 [a] | 72 [a] |
| CPU | 1.4 | 23.1 | 67.55 [a] | X |
| Difference in W | +0.3 | +19.5 | +20.6 | +4.45 |

(A) Low-End System – Generic - Single Intel i7 3770 3.4GHz 12GB DDR3 RAM

| System Type | System Power in Watts | | | |
|---|---|---|---|---|
|  | Shutdown | Idle | JTP | FPGA |
| CPU & FPGA | 7 | 172 | 209.5 [a] | 191.6 [a] |
| CPU | 6.7 | 152.5 | 184.8 [a] | X |
| Difference in W | +0.3 | +19.5 | +24.7 | +6.8 |

(B) High-End System - HP DL180 G6 - Dual Intel Xeon L5630 2.13GHz 144GB DDR3 RAM

[a.] Average power at maximum utilization

cores and threads of the CPU are active and while operating at FPGA mode only one thread is fully active (12.5 percent), while the rest of the computational intensity is offloaded to the FPGA device. At this point the difference in power consumption between the FPGA and CPU drops significantly and while still in favor of the CPU, as we can see in Table II, since the total processing speed of the CPU+FPGA far surpasses the CPU (up to 4.8X-5.3X), the total power savings goes up-to 77.8-80.7 percent on system A/B respectively.

TABLE II. SPEED UP AND POWER SAVINGS

| Algorithm Type | Speedup | System Idle Time Vs Power Savings | | |
|---|---|---|---|---|
|  |  | 0% | 10% | 20% |
| mandel | 4X | 73.4% | 72.6% | 71.8% |
| black-scholes | 4X | 73.4% | 72.6% | 71.8% |
| nbody | 4.8X | 77.8% | 77.2% | 76.5% |

(A) Low-End System – Generic - Single Intel i7 3770 3.4GHz 12GB DDR3 RAM

| Algorithm Type | Speedup | System Idle Time Vs Power Savings | | |
|---|---|---|---|---|
|  |  | 0% | 10% | 20% |
| mandel | 4.5X | 77.27% | 77.07% | 76.87% |
| black-scholes | 4.3X | 76.21% | 76.00% | 75.79% |
| nbody | 5.3X | 80.70% | 80.53% | 80.36% |

(B) High-End System - HP DL180 G6 - Dual Intel Xeon L5630 2.13GHz 144GB DDR3 RAM

Note that the best power performance results are obtained in an ideal scenario where the FPGA system is utilized at hundred percent meaning no system idle time. But since this scenario is highly unlikely in a real data center environment [8] several other utilization rates are displayed in the table. The power savings show that as long as the FPGA based system remains sufficiently utilized the benefit of using such a system outweighs the increase in idle power consumption.

### B. K-Means Algorithm

We originally experimented with twelve kernel variations with different FPGA optimization options. Several of the versions failed in synthesis because of under estimation of FPGA resource usage and others at run time because of design bugs. We eventually chose the top three performers which were hand made with custom optimizations. In addition we chose the original auto generated APARAPI version cleaned and auto optimized i.e. we only removed redundancy and inefficient parameter tagging from the APARAPI auto generated Kernel code.

*1) Speed Analysis*

Fig. 4 shows the speedup, relative to the optimized K-Means sequential Java algorithm, in time to complete 2 iterations of the K-Means MapReduce on Hadoop with N, K and D as a variable. The results show a maximum of 6.2X gain for system A and 7X for system B, for the FPGA accelerated version compared to the sequential Java version. Interestingly the auto optimized version generated by the OpenCL compiler performs worse than the sequential Java version in several cases. Another interesting note is that for different dimensions (D), different versions of the FPGA kernel perform better. This can be related to the level of loop unrolling. For example on both systems, for the 8 dimension workloads the best performance is on the 8 processing units/8 times loop unrolling version (8PU-8UL), on the other hand for 6 dimensions workloads the fastest version is the 8 processing units/3 times loop unrolling version (8PU-6UL). The performance of the JTP version is much higher than the sequential Java version since it is a highly optimized version that runs the kernel on all available CPU cores, at full utilization, but it still runs up to 44% slower and consumes up to 20% more power than the FPGA version to complete the same task.

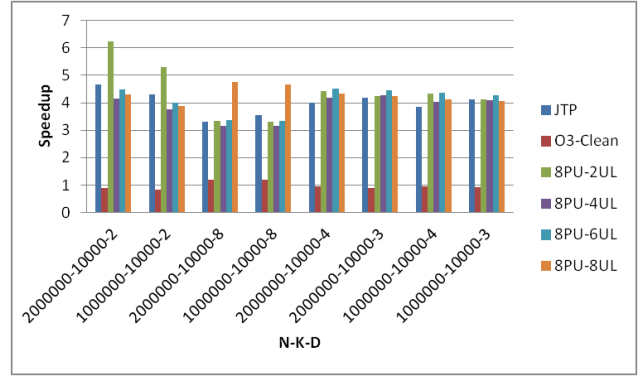

(A) Low-End System – Generic - Single Intel i7 3770 3.4GHz 12GB DDR3 RAM

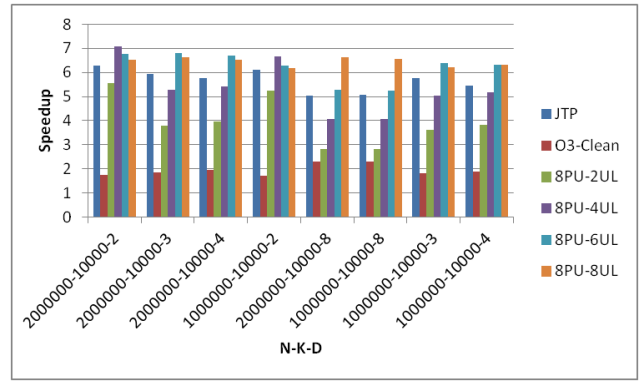

(B) High-End System - HP DL180 G6 - Dual Intel Xeon L5630 2.13GHz 144GB DDR3 RAM

Fig .4. Speedup relative to the sequential K-Means version. Different variations of the K-Means algorithm, running on different combinations of N-K-D. [Algorithm type: O3=auto optimization/ [n]PU=number of parallel processing units/ [n]UL= loop unrolling count]

*a) Data size and Calculation Complexity*

As can be seen in Fig 5, since the amount of time to setup and transfer data to an acceleration device is significant, for a low computation/data ratio, there is a negative effect to running the algorithm on the FPGA device.

The amount of floating point operations that the K-Means map kernel algorithm has to perform is in the order of N*K*D per iteration. Our experiments show that for the algorithm to run faster on the FPGA this complexity measure needs to exceed 2E+08.

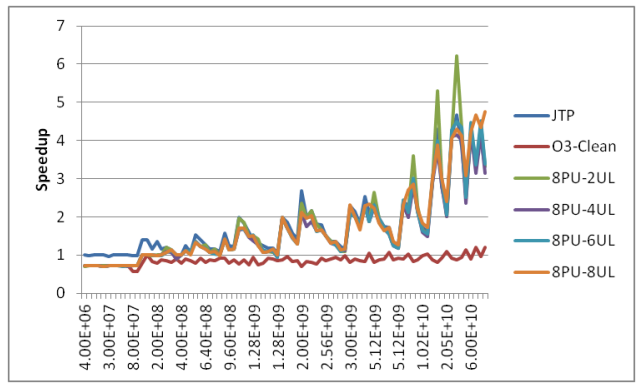

(A) Low-End System – Generic - Single Intel i7 3770 3.4GHz 12GB DDR3 RAM

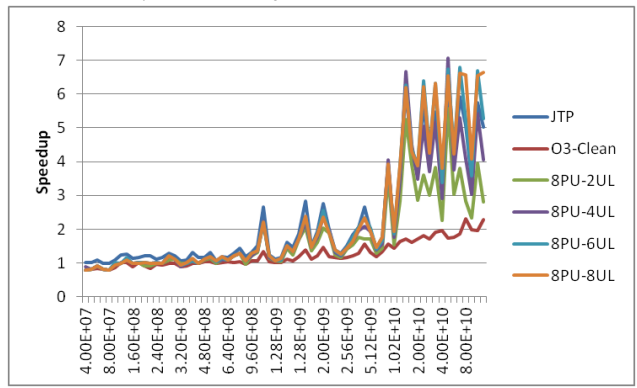

(B) High-End System - HP DL180 G6 - Dual Intel Xeon L5630 2.13GHz 144GB DDR3 RAM

Fig .5. Speedup relative to the sequential K-Means version. Different variations of the K-Means algorithm with 4E+06<=N*K*D<=8E+010.

*2) Power Analysis*

In order to assess the relative power efficiency of the different algorithm variations, we measured their average power during execution. We then calculated the average power savings in two types of cases: overall average and top ten biggest data loads, in all of which the N*K*D factor was bigger or equal to 3E+10. The results in Fig. 6 show that running the algorithm on an FPGA device can save up to 65% power on system A and up to 80% on system B when compared to the sequential version. In addition when comparing the FPGA version to the JTP version we save up to 18% and 20% on system A and B respectively.

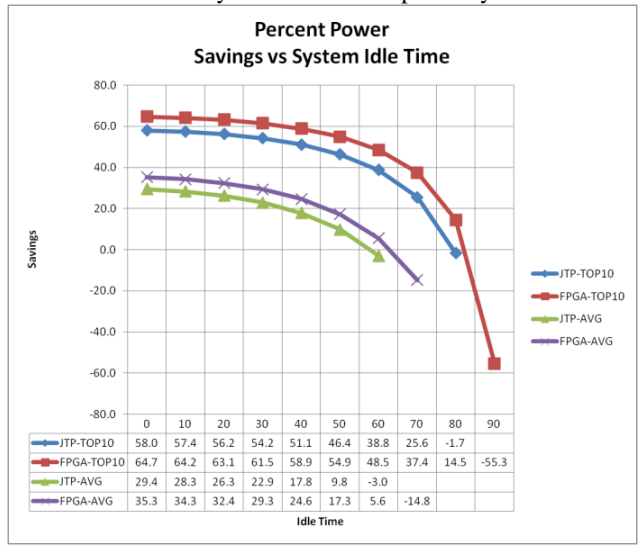

(A) Low-End System – Generic - Single Intel i7 3770 3.4GHz 12GB DDR3 RAM

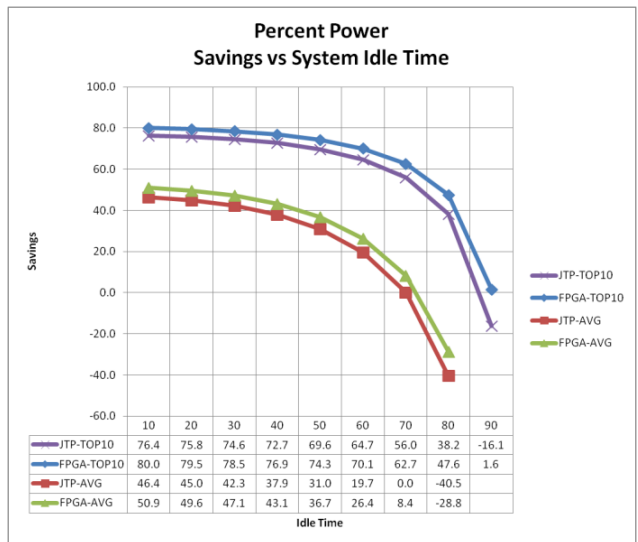

(B) High-End System - HP DL180 G6 - Dual Intel Xeon L5630 2.13GHz 144GB DDR3 RAM

Fig.6. Percent Power Savings vs Percent system Idle time relative to optimized sequential version.

*C. Performance Evaluation*

As can be observed from our power analysis on various benchmarks the host computer is responsible for the majority of system power consumption but only partially utilized during FPGA kernel execution time, hence wasting most of the energy. Low power host processors combined with external accelerator cards can offer a better solution but the performance bottleneck will remain the contemporary CPU to FPGA interface fabric that is limited by PCIE X8 bandwidth. This limits the acceleration of applications until a sufficient volume of data is available as we have experienced in our K-Means implementations. Devices with on-board ARM processors on FPGA fabric help alleviate the problems by eradicating the interfaces and sharing the same physical memory address space. Current ARM-A9 SOC devices from FPGA vendors [34, 35] are a step in the right direction, but due to the limited FPGA resources and the 4GB external memory limitation, the data-centric workloads will not scale effectively on the current generation ARM FPGA SoCs. However, the restriction of resources and memory address space could be solved by the future FPGA ARM SOCs such as Altera Stratix 10 FPGA [33] which will support quad core 64-bit ARM core with sufficient FPGA logic elements to map data-centric workloads such as K-Means.

## VII. RELATED WORK

Several high level OpenCL based programming frameworks have been developed in recent years [10, 18, and 19]. Although these libraries support OpenCL their focus is on GPU and CPU development. Higher then HDL level programming frameworks and DSLs for FPGAs have been an active research area [20, 21, 22, 28] and several open source projects exist [26, 27, 28] and continue development in the field, but they generally stop at the C/C++ level. Several commercial tool chains [23, 24] that allow a hardware flow from C/C++ to FPGA have become standard industry tools in recent years. The recent built-in Xilinx support for C/C++ and SystemC [25] and Altera's current and Xilinx planed [37] support for OpenCL in their FPGAs are more signs indicating the importance placed on high level programming of FPGAs. Existing higher then C/C++ level frameworks such as the python based MyHDL [29] and Java based MaxCompiler [30] allow high level programming of FPGAs but do not offer the programming standardization that OpenCL has to offer.

## VIII. CONCLUSIONS

On highly parallel algorithms written in Java, assuming sufficient system utilization, we find significant speedup and power savings for a system with an FPGA accelerator. On the MapReduce Hadoop framework, using the k-means algorithm we lose some of the advantages of the FPGA because of the difficulties in maximizing parallelism for a single machine. Careful analysis of computational complexity vs. data transfer overhead is needed in order to utilize the FPGA acceleration device optimally. Different OpenCL kernel versions with different optimization options should be chosen to achieve optimal performance. Custom handmade optimizations for Altera OpenCL were required for some algorithms to get good performance results, auto optimizations do not always work well.

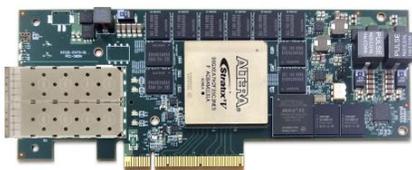

Fig 7. Nallatech PCIE-385n A7 Altera Stratix V board

## IX. FUTURE DIRECTIONS

We plan to continue our work on the framework and allow the ability to dynamically choose different target devices (FPGA/GPU/CPU etc.) at run time. In addition we would like to explore the possibility of dynamically choosing APARAPI's mode of execution and kernel versions according to an algorithm's data size and its complexity. By executing different versions of the algorithm using different modes of execution (Sequential/JTP/FPGA) on different devices we could strive to achieve the optimal power/speed solution on a per algorithm basis. In addition we plan on releasing the modification and additions we've made to APARAPI back to the open source community so it can be freely used for further research.

## X. ACKNOWLEDGMENTS

We would like to thank HP, Nallatech and Altera for their generous hardware and software donations. We would especially like to thank HP Servers for their continuous support throughout the life time of this project.

## XI. REFERENCES

[1] S. Huang, S. Xiao, and W. Feng. 2009. On the energy efficiency of graphics processing units for scientific computing. In Proceedings of the 2009 IEEE International Symposium on Parallel&Distributed Processing (IPDPS '09). IEEE Computer Society, Washington, DC, USA, 1-8.

[2] Chalamalasetti, S.; Margala, M.; Vanderbauwhede, W.; Wright, M.; Ranganathan, P., "Evaluating FPGA-acceleration for real-time unstructured search," *Performance Analysis of Systems and Software (ISPASS), 2012 IEEE International Symposium on* , vol., no., pp.200,209, 1-3 April 2012.

[3] David F. Bacon, Rodric Rabbah, and Sunil Shukla. 2013. FPGA programming for the masses. Commun. ACM 56, 4 (April 2013), 56-63.

[4] Khronos OpenCL Working Group. "OpenCL-The open standard for parallel programming of heterogeneous systems." *On line] http://www. khronos. org/opencl* (2011).

[5] Altera SDK for OpenCL. http://www.altera.com/literature/lit-opencl-sdk.jsp

[6] TIOBE Index. http://www.tiobe.com/index.php/content/paperinfo/tpci/index.html

[7] Chung, Eric S., et al. "Single-chip heterogeneous computing: Does the future include custom logic, FPGAs, and GPGPUs?." *Proceedings of the 2010 43rd Annual IEEE/ACM International Symposium on Microarchitecture*. IEEE Computer Society, 2010.

[8] Birke, Robert, Lydia Y. Chen, and Evgenia Smirni. *Data centers in the wild: A large performance study*. Technical Report RZ3820, IBM Research, 2012. http://tinyurl. com/data-centers-in-the-wild, 2012.

[9] Steve McConnell. 1993. Code Complete: A Practical Handbook of Software Construction. Microsoft Press, Redmond, WA, USA.

[10] Aparapi. API for data parallel Java, http://code.google.com/p/aparapi/.

[11] Open JDK, Project Sumatra. http://openjdk.java.net/projects/sumatra/

[12] George Kyriazis, AMD. Heterogeneous System Architecture: A Technical Review. http://developer.amd.com/wordpress/media/2012/10/hsa10.pdf

[13] Timothy Prickett Morgan. Nvidia, Continuum team up to sling Python at GPU coprocessors http://www.theregister.co.uk/2013/03/18/nvidia_continuum_pyhton_on_cuda_gpu/

[14] Jeffrey Dean and Sanjay Ghemawat. 2008. MapReduce: simplified data processing on large clusters. Commun. ACM 51, 1 (January 2008), 107-113.

[15] Apache Hadoop, http://hadoop.apache.org/core/

[16] Hartigan, John A., and Manchek A. Wong. "Algorithm AS 136: A k-means clustering algorithm." *Applied statistics* (1979): 100-108.

[17] Andreas Klöckner, Nicolas Pinto, Yunsup Lee, Bryan Catanzaro, Paul Ivanov, Ahmed Fasih, PyCUDA and PyOpenCL: A scripting-based approach to GPU run-time code generation, Parallel Computing, Volume 38, Issue 3, March 2012, Pages 157-174.

[18] JavaCL. http://code.google.com/p/javacl/

[19] Ruby-OpenCL. http://ruby-opencl.rubyforge.org/

[20] Vanderbauwhede, W.; Margala, M.; Chalamalasetti, S.R.; Purohit, S., "A C++-embedded Domain-Specific Language for programming the MORA soft processor array," *Application-specific Systems Architectures and Processors (ASAP), 2010 21st IEEE International Conference on* , vol., no., pp.141,148, 7-9 July 2010

[21] Vanderbauwhede, W.; Chalamalasetti, S.R.; Purohit, S.; Margala, M., "A few lines of code, thousands of cores: High-level FPGA programming using vector processor networks," *High Performance Computing and Simulation (HPCS), 2011 International Conference on* , vol., no., pp.461,467, 4-8 July 2011

[22] Sullivan, C.; Wilson, A.; Chappell, S., "Using C based logic synthesis to bridge the productivity gap," *Design Automation Conference, 2004. Proceedings of the ASP-DAC 2004. Asia and South Pacific* , vol., no., pp.349,354, 27-30 Jan. 2004

[23] Impulse C. http://www.impulseaccelerated.com/

[24] Catapult C. http://calypto.com/en/products/catapult/overview/

[25] Vivado Design Suite. http://www.xilinx.com/products/design-tools/vivado/

[26] A. Canis, J. Choi, M. Aldham, V. Zhang, A. Kammoona, J. Anderson, S. Brown, and T. Czajkowski. LegUp: High-level synthesis for FPGA-based processor/accelerator systems. In ACM FPGA, pages 33–36, 2011.

[27] FPGA C. http://fpgac.sourceforge.net/

[28] ROCCC. http://www.jacquardcomputing.com/roccc/

[29] MyHDL. http://www.myhdl.org/

[30] MaxCompiler.http://www.maxeler.com/products/software/maxcompiler/

[31] He, Bingsheng, et al. "Mars: a MapReduce framework on graphics processors." Proceedings of the 17th international conference on Parallel architectures and compilation techniques. ACM, 2008.

[32] Basaran, Can, and Kyoung-Don Kang. "Grex: An efficient MapReduce framework for graphics processing units." *Journal of Parallel and Distributed Computing* 73.4 (2013): 522-533.

[33] Altera ARM 64-bit (http://www.prnewswire.com/news-releases/altera-announces-quad-core-64-bit-arm-cortex-a53-for-stratix-10-socs-229678721.html)

[34] Altera ARM 32-bit cyclone. http://www.altera.com/devices/fpga/cyclone-v-fpgas/hard-processor-system/cyv-soc-hps.html

[35] Xilinx ARM 32-bit Zynq. (http://www.xilinx.com/products/silicon-devices/soc/zynq-7000/)

[36] J. Short, R. Bohn, and C. Baru, "How Much Information? 2010 Report on Enterprise Server Information," http://hmi.ucsd.edu/pdf/HMI_2010_EnterpriseReport_Jan_2011.pdf, University of California, San Diego, 2011.

[37] Xilinx OpenCL. http://www.xilinx.com/products/design-tools/all-programmable-abstractions/index.htm#software-based

[38] Hadoop K-Means Open Source Implementation. https://github.com/thomasjungblut/thomasjungblut-common/tree/master/src/de/jungblut/clustering/mapreduce